# Novel 122-type Ir-based superconductors BaIr$_2$Mi$_2$ (Mi = P and As): A density functional study

Md. Zahidur Rahaman[1]

*Department of Physics, Pabna University of Science and Technology,
Pabna-6600, Bangladesh
zahidur.physics@gmail.com*

Md. Atikur Rahman[2*]

*Department of Physics, Pabna University of Science and Technology,
Pabna-6600, Bangladesh
atik0707phy@gmail.com*
[*]Corresponding author.

## Abstract

We explore the structural, electronic, bonding, mechanical, thermodynamic and superconducting properties of two newly discovered isostructural bulk superconductors barium iridium phosphide BaIr$_2$P$_2$ ($T_c \sim 2.1$ K) and barium iridium arsenide BaIr$_2$As$_2$ ($T_c \sim 2.45$ K). The optimized structural parameters of both the compounds show good agreement with the experimentally evaluated values. The replacement of P by As mostly affects the $c$ value, whereas $a$ remains approximately the same. Metallic conductivity is observed for both the superconductors. The analysis of DOS, Mulliken atomic populations and total charge density revel a complex bonding in BaIr$_2$P$_2$ and BaIr$_2$As$_2$ with ionic, covalent and metallic characteristics. Mechanical and dynamical stability of both the phases is confirmed by analyzing the elastic constant data. According to the calculated Pugh's ratio both the intermetallics are ductile in nature. Both the superconductors demonstrate anisotropic nature whereas the elastic anisotropy of BaIr$_2$P$_2$ is higher than that of BaIr$_2$As$_2$. The hardness of BaIr$_2$P$_2$ and BaIr$_2$As$_2$ is evaluated to be 6.92 GPa and 5.02 GPa respectively indicating the relative hardness of BaIr$_2$P$_2$ than that of BaIr$_2$As$_2$ superconductor. The Debye temperature of BaIr$_2$P$_2$ and BaIr$_2$As$_2$ has been calculated by using the elastic constant data to be 293.06 K and 258.47 K respectively. Finally the electronic specific heat coefficient, electron-phonon coupling constant, coulomb pseudo potential and superconducting critical temperature have been evaluated for both the compounds. We found weakly and moderately coupled bulk superconductivity in BaIr$_2$P$_2$ and BaIr$_2$As$_2$ with transition temperature 0.0057 K and 6 K respectively.

## I. Introduction

The discovery of superconductivity in Fe-based compounds gives a clear spur to the search for new superconductors with high superconducting critical temperature [1]. The material combination of iron pnictide is very promising for practical discovery of superconductors with high transition temperature. The combination of ternary 122-type materials is alike as that of so called iron pnictide and hence in this structure type the substitution of iron by any other transition metals provides a fresh route for the exploration of high $T_c$ superconducting materials.

On the other hand the rare-earth compounds with AM$_2$X$_2$ type structure (where, A = lanthanide element or any alkaline earth element; M = transition metal; X = P, As, Ge or Si) have attracted a large scientific community with their many attractive properties including heavy fermion behavior, superconductivity, exotic magnetic order, mixed valency and valence fluctuation [2]. Since one possible route to explore high $T_c$ superconductors is the substitution of Fe by other transition elements in the AM$_2$X$_2$ type structure, ThCr$_2$Si$_2$ structured compounds can be the great choice with more than seven hundred representatives. ThCr$_2$Si$_2$ structure type was first reported in 1965 by Ban and Sikirica [3]. Recently after the discovery of $(Ba_{0.6}K_{0.4})Fe_2As_2$ superconductor [4] ($T_c$ = 38K) the $ThCr_2Si_2$ structure type has become a family of large interest to the researchers. Besides the Fe-based compounds the Pd, Ni and Pt-based compounds with $ThCr_2Si_2$ structure type [5-7] have been discovered recently with $T_c$ up to 23K which shade a light to form a family of novel high $T_c$ superconductors.

In Fe-based compounds the $AFe_2As_2$ (A = Sr, Ca, Ba, etc.) structure type is free of oxygen and exhibits metallic conductivity [8-10]. These compounds have been investigated extensively to understand their superconducting mechanism [11, 12]. Due to the presence of Ir in the transition metal group Fe can be replaced by Ir and alike superconductivity can be expected in Ir-based compounds. Among Ir-based superconductors $BaIr_2P_2$ is investigated extensively. This compound is a conventional electron-phonon superconductor with transition temperature up to 2.1K [13-15]. Recently Forster *et al*. reported large effective mass enhancements in $BaIr_2P_2$ by using de Haas-Van-Alphen quantum oscillation study [16]. Berry *et al*. proved the presence of bulk weak coupling superconductivity in Barium iridium phosphide [15]. He also suggests that the lack of interlayer bonding favors superconductivity in $BaIr_2P_2$ [15]. D. Billington *el al*. calculates the electron phonon coupling constant of $BaIr_2P_2$ intermetallic as 0.52 according to which $BaIr_2P_2$ is a moderate coupling superconductor [17]. So the results obtaining from the investigation of Berry *et al*. and D. Billington *el al*. show some contradiction and it will be discussed in this present letter. Superconductivity is also found in other Ir-based compounds such as ScIrP, LaIrAs, LaIrP, $CeIrSi_3$ and $IrSe_2$ [18-22]. The ternary intermetallic $SrIr_2As_2$ also shows superconductivity with transition temperature as approximately 2.9K [23]. The ternary barium iridium arsenide $BaIr_2As_2$ ($ThCr_2Si_2$ type structure) is synthesized by Xiao-Chuan *et al.* in 2016 [24]. He with his co-workers suggested the type-II bulk superconductivity with superconducting transition temperature as 2.45K in $BaIr_2As_2$.

In spite of being a considerable advancement in understanding about the electronic and superconducting properties of barium iridium phosphide, the detailed theoretical works on the mechanical and thermodynamic properties of this superconductor is still absence in literature. On the other hand, no theoretical report is available in literature about the detailed physical properties of barium iridium arsenide. In this present letter we therefore decide to study the detailed physical and superconducting properties of these novel ternary superconductors $BaIr_2P_2$ and $BaIr_2As_2$ by using the DFT based CASTEP code to have a better understanding about the physical and superconducting nature of these intermetallics. Finally a thorough comparison among the characteristics of these two superconductors has also been made with proper discussion.

## II. Computational details

Ab-initio calculations based on DFT (Density Functional Theory) were carried out in the CASTEP computer code by using the generalized gradient approximation (GGA) with the Perdew-Burke-Ernzerhof (PBE) exchange correlation functional [27]. The plane wave cut-off energy was set to 310 eV for both the compounds and Monkhorst-Pack grid with 84 irreducible k-points (6 × 6 × 9) was used to perform Brillouin-zone sampling. Pseudo atomic calculations were performed for P-$3s^2$ $3p^3$, Ba-$5s^2$ $5p^6$ $6s^2$ and Ir-$5d^7$ $6s^2$ in case of $BaIr_2P_2$ superconductor and As-$4s^2$ $4p^3$, Ba-$5s^2$ $5p^6$ $6s^2$ and Ir-$5d^7$ $6s^2$ in case of $BaIr_2As_2$ superconductor. BFGS (Broyden-Fletcher-Goldfarb-Shanno) minimization algorithm was used to optimize the geometric structure of both the superconductors [28]. Voigt-Reuss-Hill approximation was used to estimate the single and polycrystalline elastic constants [29].

## III. Results and discussion

### A. Structural properties

At ambient temperature and pressure, BaIr$_2$Mi$_2$ (Mi = P and As) adopts a tetragonal layered ThCr$_2$Si$_2$ type structure with the space group of *I4/mmm* (139) [15, 24]. The optimized crystal structure of BaIr$_2$Mi$_2$ is shown in Fig. 1. The unit cell contains two formula units with ten atoms. There is one formula unit per primitive cell with five atoms. In a unit cell of BaIr$_2$Mi$_2$ (Mi = P and As) Ba atoms sit at 2*a* (0, 0, 0), Ir atoms sit at 4*d* (0, 0.5, 0.25) and Mi atoms sit at 4*e* (0, 0, 0.3593) Wyckoff position [30]. The unit cell dimensions including equilibrium lattice parameters for tetragonal phase $a_0$ and $c_0$, bulk modulus $B_0$ and unit cell volume $V_0$ of BaIr$_2$P$_2$ and BaIr$_2$As$_2$ intermetallics at ambient temperature are listed in Table 1 with the experimentally evaluated values. From Table 1 one can see that the evaluated unit cell parameters show good consistence with the experimental data bearing the reliability of this present study. The existing minor deviation of the calculated values from the experimental ones may cause by the temperature dependency of cell parameters and GGA process [31]. One can notice from Table 1 that the substitution of P by As mostly influences the *c* value. The bulk modulus is also increased from 115.60 GPa to 149.52 GPa when P is replaced by As.

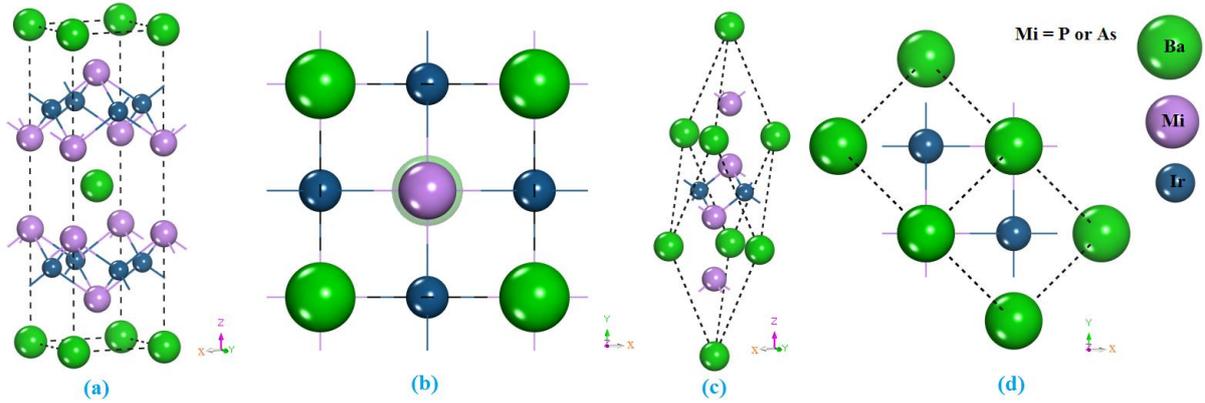

**Fig. 1.** The crystal structures of BaIr$_2$Mi$_2$ (Mi = P and As). (a) Conventional unit cell (three dimensional); (b) Conventional unit cell (two dimensional); (c) Primitive cell (three dimensional); (d) Primitive cell (two dimensional).

**Table 1.** Unit cell dimensions of ThCr$_2$Si$_2$-type BaIr$_2$P$_2$ and BaIr$_2$As$_2$ superconductors.

| Properties | BaIr$_2$P$_2$ | | BaIr$_2$As$_2$ | |
|---|---|---|---|---|
| | This study | Expt. [15] | This study | Expt. [24] |
| $a_0$ (Å) | 3.983 | 3.946 | 4.087 | 4.052 |
| $c_0$ (Å) | 12.712 | 12.559 | 13.094 | 12.787 |
| $c_0/a_0$ | 3.191 | 3.182 | 3.203 | 3.155 |
| $V_0$ (Å$^3$) | 201.66 | 195.55 | 218.71 | 209.94 |
| $B_0$ (GPa) | 115.60 | - | 112.52 | - |

## B. Single and polycrystalline elastic properties

In order to get a deep insight about the nature of force in crystal it is important to study the elastic properties of solid. Through the detailed investigation into the elastic properties we can have a lucid idea about the dynamical behavior of materials. Various important properties of material such as ductility, anisotropy, stiffness, brittleness and stability can be derived from the elastic constants data [32]. Hence in this section a thorough investigation into the mechanical nature of $BaIr_2P_2$ and $BaIr_2As_2$ superconductors has been done with proper discussion and comparison.

The elastic constants of $BaIr_2Mi_2$ (Mi = P and As) are estimated in order to the Hook's law by linear fitting of the calculated stress-strain function [33]. A crystal with the tetragonal phase belongs to six independent elastic constants ($C_{11}$, $C_{12}$, $C_{13}$, $C_{33}$, $C_{44}$ and $C_{66}$). The estimated elastic constants of $BaIr_2P_2$ and $BaIr_2As_2$ superconductors are illustrated in Table 2. According to the stability criteria [34] of tetragonal phase (see Eq.1) both the compounds under consideration have good stability in nature.

$$\left.\begin{array}{l} C_{11} > 0,\ C_{33} > 0,\ C_{66} > 0,\ C_{44} > 0 \\ C_{11} + C_{33} - 2C_{13} > 0,\ C_{11} - C_{12} > 0 \\ 2(C_{11} + C_{12}) + 4C_{13} + C_{33} > 0 \end{array}\right\} \quad (1)$$

It can be seen from Table 2 that the value of $C_{11}$ is much larger than the value of $C_{33}$ for both superconductors implying that the bonding strength between (001) and (100) planes are not the same [35]. It also implies that the incompressibility along [001] direction is weaker than that of [100] direction [36]. Again for both the compounds [100](010) shear is harder than the [100](001) shear since $C_{66} > C_{44}$ [36]. The evaluated elastic constants of $BaIr_2P_2$ and $BaIr_2As_2$ superconductors show reasonably good consistent with the elastic constants of $LaRu_2P_2$ and $LaRu_2As_2$ superconductors [2] showing the reliability of this present investigation. However, the incompressibility and the shear resistance of $BaIr_2As_2$ are weaker than that of $BaIr_2P_2$ superconductor as most of the elastic constants of $BaIr_2As_2$ are smaller than the $BaIr_2P_2$ intermetallic [2].

**Table 2.** The evaluated elastic constants $C_{ij}$ (in GPa) of $BaIr_2P_2$ and $BaIr_2As_2$ superconductors.

| | Elastic constants | | | | | |
|---|---|---|---|---|---|---|
| **Compounds** | $C_{11}$ | $C_{12}$ | $C_{13}$ | $C_{33}$ | $C_{44}$ | $C_{66}$ |
| **$BaIr_2P_2$** | 241.36 | 106.68 | 47.12 | 119.82 | 30.05 | 124.97 |
| **$BaIr_2As_2$** | 204.67 | 90.62 | 67.34 | 111.14 | 34.79 | 95.06 |

The bulk modulus $B$, Young's modulus $E$, shear modulus $G$ and Poisson's ratio $v$ of both the compounds are determined By using the Voigt-Reuss-Hill (VRH) averaging scheme [37]. In this approximation the bulk and shear moduli are given as follows:

$$B_V = \frac{2C_{11} + 2C_{12} + C_{33} + 4C_{13}}{9} \quad (2)$$

$$B_R = \frac{C^2}{M} \tag{3}$$

$$G_V = \frac{M + 3C_{11} - 3C_{12} + 12C_{44} + 6C_{66}}{30} \tag{4}$$

$$G_R = \frac{15}{\left[\frac{18B_V}{C^2} + \frac{6}{(C_{11} - C_{12})} + \frac{6}{C_{44}} + \frac{3}{C_{66}}\right]} \tag{5}$$

Where, $M$ and $C^2$ can be written as,

$$M = C_{11} + C_{12} + 2C_{33} - 4C_{13}$$

And

$$C^2 = (C_{11} + C_{12})C_{33} - 2C_{13}^2$$

Now to determine $B$ and $G$ an arithmetic mean were taken by Hill are given as follows,

$$B = \frac{1}{2}(B_R + B_v) \tag{6}$$

$$G = \frac{1}{2}(G_v + G_R) \tag{7}$$

The estimated values of $B$ and $G$ by using Eq.6 and Eq.7 are illustrated in Table 3. One can notice from Table 3 that both the compounds possess low value of bulk modulus indicating the soft nature of these intermetallics. It is easier to change the volume of $BaIr_2As_2$ superconductor under applied pressure than that of $BaIr_2P_2$ since the $B$ value of $BaIr_2P_2$ is larger than the $BaIr_2As_2$ superconductor [38]. The shear modulus of $BaIr_2P_2$ is larger than that of $BaIr_2As_2$ superconductor showing higher shear resistance and strong covalent bond in $BaIr_2P_2$ [39].

Now the Young's modulus ($E$) and the Poisson's ratio ($v$) can be calculated as follows by using the $B$ and $G$ values,

$$E = \frac{9GB}{3B + G} \tag{8}$$

$$v = \frac{3B - 2G}{2(3B + G)} \tag{9}$$

The evaluated Young's modulus of $BaIr_2As_2$ is smaller than $BaIr_2P_2$ (Table 3) indicating that the $BaIr_2As_2$ superconductor is less stiff than that of $BaIr_2P_2$ [40]. To understand the nature of bonding force in a solid the Poisson's ratio of both the compounds has been evaluated and tabulated in Table 3. From Table 3 it is evident that the value of $v$ lies between 0.25 and 0.50 indicating that the force exist in $BaIr_2Mi_2$ is central [41]. Pugh's ratio ($B/G$) is particularly useful to predict the ductility and brittleness of a crystal [40]. From Table 3 one can see that both the compounds demonstrate ductile nature since $B/G > 1.75$ represents ductile nature of a crystal.

**Table 3.** Evaluated polycrysttalline bulk modulus $B$ (GPa), shear modulus $G$ (GPa), Young's modulus $E$ (GPa), $B/G$ value, Poisson's ratio $v$, elastic anisotropy $A^U$ and Vickers hardness $H_v$ (GPa) of $BaIr_2P_2$ and $BaIr_2As_2$ superconductors.

| | Polycrystalline elastic properties | | | | | | |
|---|---|---|---|---|---|---|---|
| Compounds | $B$ | $G$ | $E$ | $B/G$ | $v$ | $A^U$ | $H_v$ |
| $BaIr_2P_2$ | 102.46 | 55.04 | 140.04 | 1.86 | 0.27 | 2.09 | 6.92 |
| $BaIr_2As_2$ | 101.78 | 48.75 | 126.11 | 2.08 | 0.29 | 0.98 | 5.02 |

The anisotropic nature of a solid can be calculated by using the following relation [42],

$$A^U = \frac{5G_V}{G_R} + \frac{B_V}{B_R} - 6 \tag{10}$$

$A^U = 0$ indicates completely isotropic crystal and the deviation from this value shows the degree of anisotropy in the crystal. From Table 3 it is clear that the $BaIr_2P_2$ superconductor is more anisotropic than that of $BaIr_2As_2$ superconductor. Chung and Buessen suggests two new relations [43] to determine the anisotropy indexes of bulk and shear moduli given as follows,

$$A_B = \frac{(B_V - B_R)}{(B_V + B_R)} \tag{11}$$

$$A_G = \frac{(G_V - G_R)}{(G_V + G_R)} \tag{12}$$

Where, for a completely isotropic nature of a compound $A_B = A_G = 0$ and $A_B = A_G = 1$ demonstrates the maximum anisotropy in a crystal. For $BaIr_2P_2$ $A_B = 0.09$ and $A_G = 0.15$ and in case of $BaIr_2As_2$ $A_B = 0.06$ and $A_G = 0.08$ showing the anisotropic nature of both the compounds. Chen et al. proposed a new theoretical model to calculate the hardness of a solid which is given as [43],

$$H_V = 2(K^2 G)^{0.585} - 3 \tag{13}$$

It is evident from Table 3 that $BaIr_2As_2$ is relatively soft compound compared to $BaIr_2P_2$ superconductor.

## C. Electronic properties and chemical bonding

In this section we have presented and discussed the band structure, total and partial density of states, total charge density and Mulliken atomic populations of $BaIr_2P_2$ and $BaIr_2As_2$ superconductors to have a lucid insight into the electronic characteristics and chemical bonding of these compounds. The band structure diagrams for both the compounds are depicted in Fig. 2. It is evident from Fig. 2 that the valence bands and conduction bands are overlapped at Fermi level ($E_F$) for both the intermetallics, showing the metallic feature of $BaIr_2P_2$ and $BaIr_2As_2$. The metallic conductivity of $BaIr_2P_2$ and $BaIr_2As_2$ indicates that both the intermetallics can be superconductor [32].

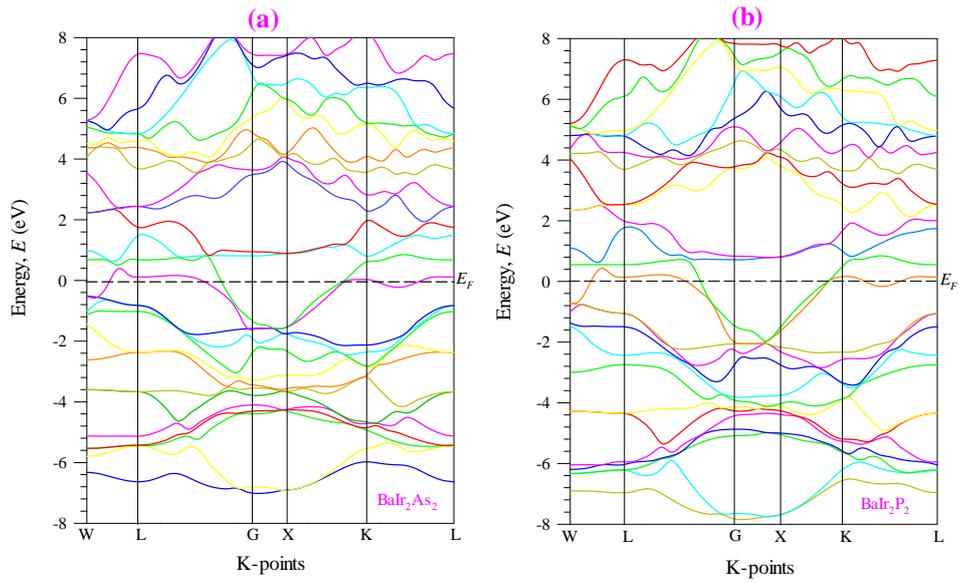

**Fig. 2.** Electronic band structure of (a) BaIr$_2$As$_2$ and (b) BaIr$_2$P$_2$ ternary intermetallics.

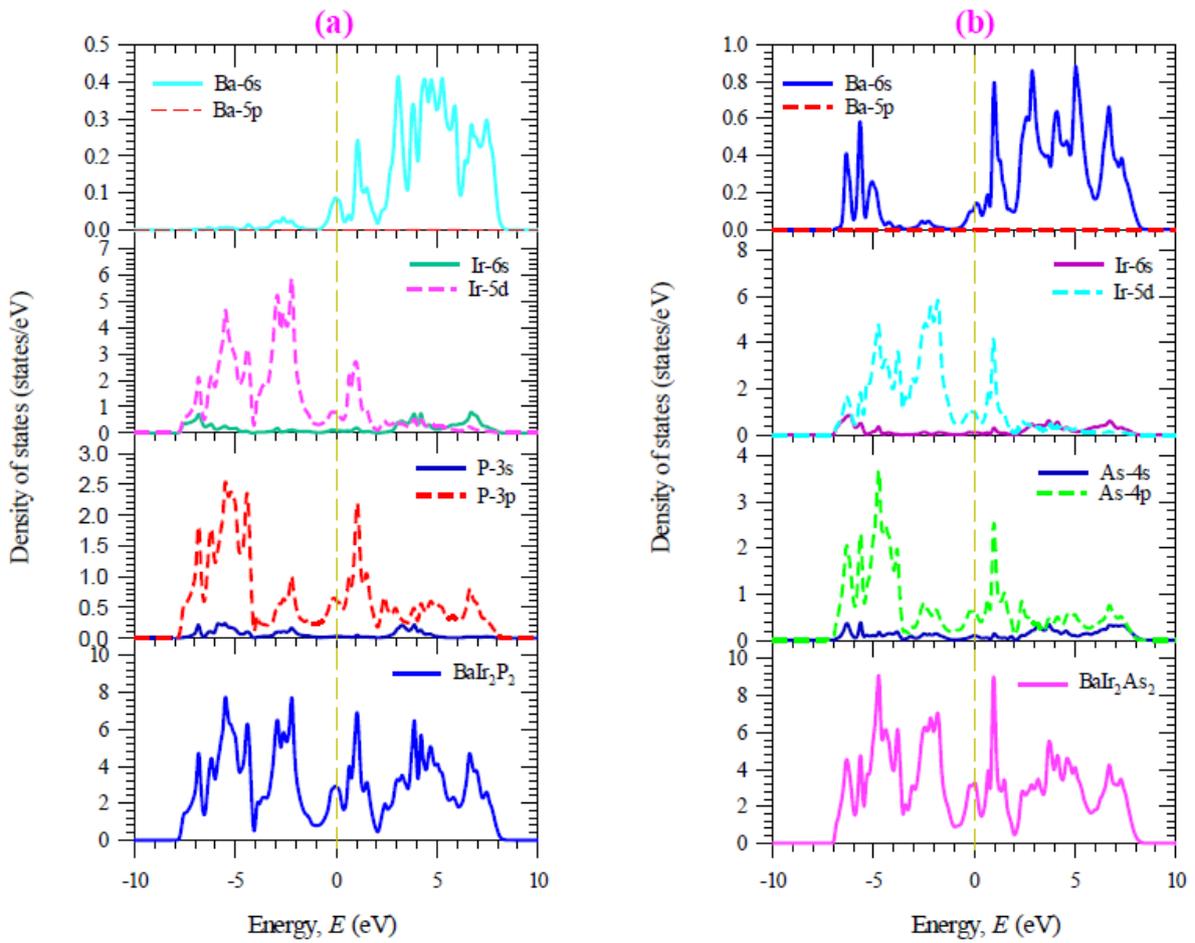

**Fig. 3.** The density of states (partial and total) of (a) BaIr$_2$P$_2$ and (b) BaIr$_2$As$_2$ superconductors.

The density of states (total and partial) of BaIr$_2$P$_2$ and BaIr$_2$As$_2$ intermetallics is depicted in Fig. 3. According to Fig. 3, the total density of states (TDOS) of both the compounds is composed of four main peaks. The first peak in the valence band lies between -7.84 eV and -4.13 eV in case of BaIr$_2$P$_2$ and from -7 eV to -3.58 eV in case of BaIr$_2$As$_2$ superconductor. In case of BaIr$_2$P$_2$ superconductor, Ir-5d and P-3p states contribute the most to constitute the first peak whereas the contribution of Ir-5d states is dominant. Similarly in BaIr$_2$As$_2$, the contribution of Ir-5d states is dominant whereas Ba-6s, Ir-5d and As-4p states contribute to constitute the first peak of BaIr$_2$As$_2$. The second peak lies between -4.13 eV and -1.06 eV in case of BaIr$_2$P$_2$ and from -3.58 to -0.89 eV in case of BaIr$_2$As$_2$ intermetallic. This peak is dominated by Ir-5d band in both the compounds. The third peak lies in the conduction band from 0.31 eV to 1.98 eV and 0.34 eV to 1.34 eV for BaIr$_2$P$_2$ and BaIr$_2$As$_2$ superconductors respectively. This peak is contributed by Ba-6s, Ir-5d, P-3p and Ba-6s, Ir-5d, As-4p states for BaIr$_2$P$_2$ and BaIr$_2$As$_2$ superconductors respectively in which the Ir-5d state is dominant for both the compounds. The fourth peak lies in 1.98 eV and 8.10 eV in case of BaIr$_2$P$_2$ and 2.02 eV and 8.06 eV in case of BaIr$_2$As$_2$ superconductor. This peak is dominated by Ba-6s state for both the compounds. We observe clear coincidence between the P-3p and Ir-5d states in case of BaIr$_2$P$_2$ and As-4p and Ir-5d states in case of BaIr$_2$As$_2$ superconductors, which suggests the covalent nature of P-Ir and As-Ir bonds in BaIr$_2$P$_2$ and BaIr$_2$As$_2$ compounds respectively [44]. This is a common feature of ThCr$_2$Si$_2$ type compounds [45]. Moreover, the first peak of valence band revels that the P(3p)-Ir(5d) bonds in BaIr$_2$P$_2$ are set at higher energy range than the As(4p)-Ir(5d) bonds in BaIr$_2$As$_2$ indicating that the binding energy of As-Ir bonds in BaIr$_2$As$_2$ are higher and stronger than that of P-Ir bonds in BaIr$_2$P$_2$ [46]. The calculated DOS at $E_F$ is 2.88 and 3.28 states/eV fu for BaIr$_2$P$_2$ and BaIr$_2$As$_2$ compounds respectively mainly contributed by Ir-5d states electrons.

For gaining further insight into the bonding nature of these two superconductors, the Mulliken atomic population [47-49] is evaluated and listed in Table 4. The study of the Mulliken population provides useful information about the bond overlap and the transfer of charge among different atoms in a compound. It is evident from Table 4 that P and Ir atoms contain the negative charges while Ba atoms contain the positive charges in case of BaIr$_2$P$_2$ superconductor indicating the transferring of charge from Ba to P and Ir atoms. Similarly charge transfers from Ba and As to Ir atoms in case of BaIr$_2$As$_2$ superconductor. These results imply that both the superconductors have some ionic feature. It is also evident from Table 4 that the population of P-Ir bonds is positive showing the covalent character of P-Ir bonds in BaIr$_2$P$_2$. This result agrees well with the result having from the DOS analysis. In BaIr$_2$As$_2$ the As-Ir bonds have some negative value indicating the ionic character of these bonds. This result shows direct contradiction with the common feature of ThCr$_2$Si$_2$ structured compounds. For clearing this contradiction we have also investigated the charge density map for both the compounds. For more comprehensive understanding the ionicity of bond is calculated by using the below equation [50],

$$f_h = 1 - e^{-|P_c - P|/P} \qquad (14)$$

Where, $P$ stands for bond overlap population and $P_c$ implies the population in pure covalent compound. For a pure covalent compound $P_c = 1$. According to the calculated value of $f_h$ P-Ir bonds have covalent nature since the value greater than zero indicates the increasing level of covalency. Whereas the Ir-Ir bonds in BaIr$_2$As$_2$ have low level of covalency and comparatively high level of ionicity since the $f_h$ value is far from the condition of a pure covalent crystal ($f_h = 1$).

**Table 4.** Mulliken atomic populations of $BaIr_2P_2$ and $BaIr_2As_2$ superconductors.

| | Species | s | p | d | Total | Charge | Bond | Population | $f_h$ | Lengths (Å) |
|---|---|---|---|---|---|---|---|---|---|---|
| **$BaIr_2P_2$** | P | 1.81 | 3.49 | 0.00 | 5.30 | -0.30 | | | | |
| | Ba | 1.71 | 6.04 | 1.07 | 8.83 | 1.17 | P-Ir | 0.60 | 0.48 | 2.370 |
| | Ir | 0.82 | 0.41 | 8.05 | 9.29 | -0.29 | | | | |
| **$BaIr_2As_2$** | As | 0.72 | 3.53 | 0.00 | 4.25 | 0.75 | As-Ir | -1.05 | - | 2.477 |
| | Ba | 2.72 | 6.04 | 1.13 | 9.89 | 0.11 | Ir-Ir | 0.75 | 0.28 | 2.890 |
| | Ir | 0.90 | 0.87 | 8.03 | 9.81 | -0.81 | | | | |

In order to get clear insight into the bonding the total charge density for both the compounds in the direction of (100) plane is investigated as shown in Fig. 4. A covalent feature is appeared of P-Ir and As-Ir bonds in $BaIr_2P_2$ and $BaIr_2As_2$ respectively since a clear overlapping of electron (charge) distribution is appeared among these atoms. There is no overlapping of charge distribution among Ir atoms in both the compounds showing the ionic character of Ir-Ir bonds. The ionic character is the consequence of the metallic character [51] showing the metallic feature of Ir-Ir bonds.

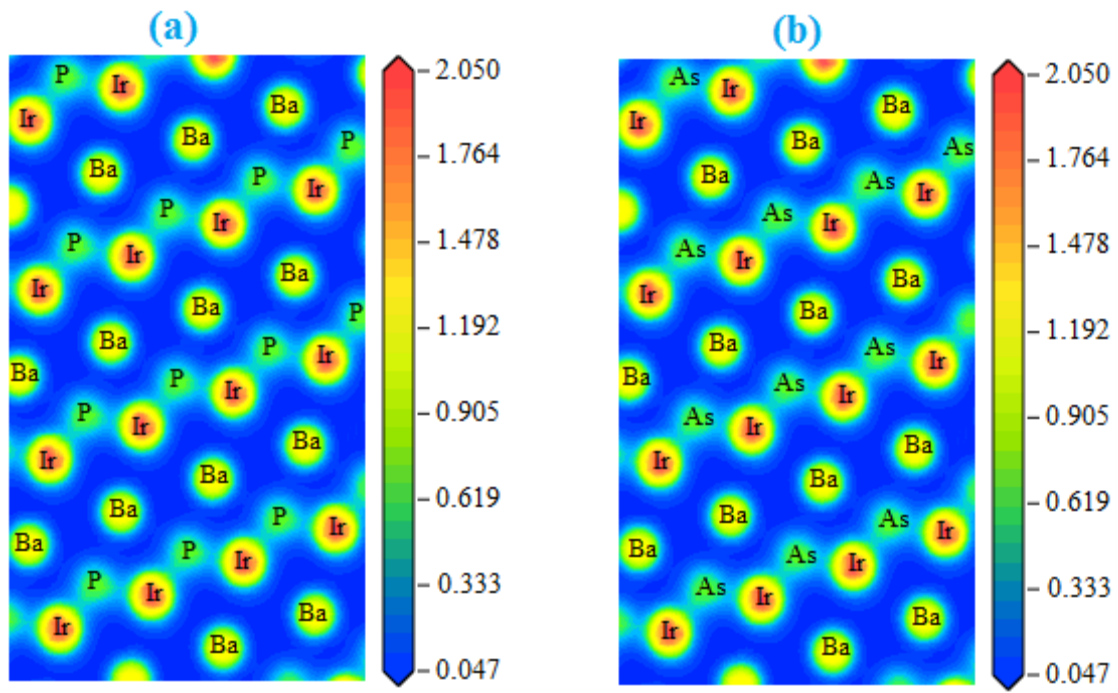

**Fig. 4.** The total charge density of (a) $BaIr_2P_2$ and (b) $BaIr_2As_2$ superconductors along (100) plane.

Overall, the detailed study of the DOS, Mulliken atomic population, and total charge density of $BaIr_2P_2$ and $BaIr_2As_2$ superconductors indicate the existence of all the ionic, covalent and metallic bonds in these compounds which is the common feature of $ThCr_2Si_2$ structured compounds.

## D. Thermodynamic properties

The study of the thermodynamic properties of crystal is crucial which can lead to deep insight into the many solid-state phenomena. In this section, we have investigated the Debye temperature, thermal conductivity, melting temperature and Dulong-Petit limit of $BaIr_2P_2$ and $BaIr_2As_2$ superconductors. The calculated values are tabulated in Table 5 and Table 6.

The Debye temperature ($\Theta_D$) is an important thermodynamic quantity which is directly related to the many thermal properties such as melting point of solid, thermal expansion, specific heat of solid etc. It is defined as the temperature which relates with the highest normal mode of vibration of solid. The Debye temperature of $BaIr_2P_2$ and $BaIr_2As_2$ superconductors has been calculated by using the evaluated elastic constants data following the below equation [52],

$$\theta_D = \frac{h}{k_B}\left(\frac{3N}{4\pi V}\right)^{\frac{1}{3}} \times v_m \qquad (15)$$

Where, $V_m$ is the average sound velocity which is given by,

$$v_m = \left[\frac{1}{3}\left(\frac{2}{v_t^3} + \frac{1}{v_l^3}\right)\right]^{-\frac{1}{3}} \qquad (16)$$

Where, $V_t$ and $V_l$ are the transverse and longitudinal wave velocity respectively.

$$v_l = \left(\frac{3B + 4G}{3\rho}\right)^{\frac{1}{2}} \qquad (17)$$

And

$$v_t = \left(\frac{G}{\rho}\right)^{\frac{1}{2}} \qquad (18)$$

The calculated $V_m$, $V_t$, $V_l$, $\rho$ and $\Theta_D$ of $BaIr_2P_2$ and $BaIr_2As_2$ intermetallics are tabulated in Table 5 with the available experimental values. As shown in Table 5 our evaluated values show good agreement with the experimentally evaluated values bearing the reliability of our present investigation.

**Table 5.** The calculated density $\rho$ (in gm/cm$^3$), transverse ($V_t$), longitudinal ($V_l$), and average sound velocity $V_m$ (m/s) and Debye temperature $\Theta_D$ (K) of $BaIr_2P_2$ and $BaIr_2As_2$ superconductors.

| Compounds | $\rho$ | $V_t$ | $V_l$ | $V_m$ | $\Theta_D$ | Remarks |
|---|---|---|---|---|---|---|
| **$BaIr_2P_2$** | 9.60 | 2394.43 | 4279.87 | 2669.30 | 293.06 | This study |
| | - | - | - | - | 292 | Expt. [1] |
| **$BaIr_2As_2$** | 10.62 | 2142.52 | 3962.86 | 2392.72 | 258.47 | This study |
| | - | - | - | - | 202 | Expt. [24] |

The ability to conduct heat energy from one part to the other part of a solid is generally defined as the thermal conductivity of solid. The minimum thermal conductivity, $K_{min}$ is an important thermal parameter which is directly related to the high temperature applications of solid [53]. When the temperature of a solid is gradually increased the conductivity of the solid is then gradually decreased to a certain limit which is usually referred to as the minimum thermal conductivity of that solid [53]. In this present study the minimum thermal conductivity of $BaIr_2P_2$ and $BaIr_2As_2$ superconductors has been calculated by using the following expression [54],

$$K_{min} = K_B v_m \left(\frac{M}{n\rho N_A}\right)^{-2/3} \tag{19}$$

Where, $K_B$ is the Boltzmann constant, $v_m$ is the average sound velocity, $M$ is defined as the molecular mass, $n$ stands for the number of atoms per molecule and $N_A$ is defined as the Avogadro's number. The evaluated $K_{min}$ of $BaIr_2P_2$ and $BaIr_2As_2$ intermetallics by using Eq. 19 is listed in Table 6. As shown in Table 6 the calculated value of $K_{min}$ for $BaIr_2P_2$ is somewhat larger than that of $BaIr_2As_2$ superconductor. However, at ambient condition both the compounds show comparatively low thermal conductivity.

**Table 6.** The evaluated minimum thermal conductivity, $K_{min}$ (in Wm$^{-1}$K$^{-1}$), melting temperature, $T_m$ (K) and the Dulong-Petit limit (J/mole.K) of $BaIr_2P_2$ and $BaIr_2As_2$ superconductors.

| Compounds | $K_{min}$ | $T_m$ | Dulong-Petit limit |
|---|---|---|---|
| $BaIr_2P_2$ | 0.49 | 1257.81 | 124.67 |
| $BaIr_2As_2$ | 0.43 | 1134.72 | |

The melting temperature of tetragonal metals can be evaluated by using Eq. 20 [55],

$$T_m = 354 + \frac{4.5\,(2C_{11} + C_{33})}{3} \tag{20}$$

The evaluated melting temperature of $BaIr_2P_2$ and $BaIr_2As_2$ intermetallics is listed in Table 6. From Table 6 it is evident that the melting temperature of $BaIr_2P_2$ is larger than that of $BaIr_2As_2$ superconductor. At high temperature the anharmonic effect of the specific heat at constant volume is suppressed and close to a limit known as Dulong-Petit limit [56]. The Dulong-Petit limit of solid can be estimated by the following relation [56],

$$Dulong - Petit\ limit = 3nN_A K_B \tag{21}$$

Where, $N_A$ is the Avogadro's constant and $K_B$ represents the Boltzmann constant. The estimated Dulong-Petit limit of both the compounds is tabulated in Table 6.

**E. Superconducting properties**

Based on the strong-coupled theory of superconductivity W.L. McMillan shows a relation among the superconducting critical temperature ($T_c$), Coulomb coupling constant and the electron-phonon coupling constant ($\lambda$) which shows good accuracy in case of real metals have lowest order of an expansion parameter $\frac{\hbar\omega_{ph}}{E_F} \sim 10^{-2} - 10^{-3}$ [57]. The relation derived by McMillan does not consider the anisotropy of the energy gap but takes into account the persistent spin fluctuations which are important for the nearly ferromagnetic materials. This theoretical relation for obtaining $T_c$ directly is given as follows [57],

$$T_c = \frac{\theta_D}{1.45} e^{-\frac{1.04\,(1+\lambda)}{\lambda - \mu^*(1+0.62\lambda)}} \qquad (22)$$

Where, $\theta_D$, $\lambda$ and $\mu^*$ are defined as the Debye temperature, electron-phonon coupling constant and coulomb pseudo potential respectively. The detailed procedures of obtaining these parameters are discussed elsewhere from Eq. 1 to Eq. 4 [58]. The estimated superconducting parameters of BaIr$_2$P$_2$ and BaIr$_2$As$_2$ intermetallics by using Eq. 1 to Eq. 4 of Ref. 58 are listed in Table 7 with the available experimental values.

**Table 7.** Evaluated density of states at Fermi level $N\,(E_F)$ (states eV$^{-1}$ fu$^{-1}$), specific heat coefficient $\gamma_{bs}$ (calculated by using band structure data) and $\gamma$ (mJ/ K$^2$ mol), electron-phonon coupling constant $\lambda$, the coulomb pseudo potential $\mu^*$ and transition temperature $T_c$ (K) of BaIr$_2$P$_2$ and BaIr$_2$As$_2$ ternary intermetallics.

| Compounds | Remarks | $N(E_F)$ | $\gamma_{bs}$ | $\gamma$ | $\lambda$ | $\mu^*$ | $T_c$ |
|---|---|---|---|---|---|---|---|
| **BaIr$_2$P$_2$** | This Cal. | 2.88 | 6.77 | 9.27 | 0.37 | 0.19 | 0.0057 |
| | Exp. [15] | - | - | 9.30 | - | - | 2.10 |
| | [1] | - | - | 6.86 | - | - | - |
| | Other Cal.[17] | 3.01 | 7.07 | - | 0.52 | - | 2.05 |
| | [25] | - | 7.21 | - | - | - | - |
| | [26] | 2.65 | - | 9.65 | 0.55 | - | 1.97 |
| **BaIr$_2$As$_2$** | This Cal. | 3.28 | 7.72 | 14.59 | 0.89 | 0.20 | 6.0 |
| | Exp. [24] | - | - | 14.60 | - | - | 2.45 |
| | Other Cal. | - | - | - | - | - | - |

As shown in Table 7, our evaluated value of $N(E_F)$ shows good consistent with the other theoretical letters in case of BaIr$_2$P$_2$ superconductor. The estimated value of electronic specific heat coefficient ($\gamma$) also agrees with the value obtained in Ref. 15 but shows contradiction with Ref. 1. However, based on this present study which supports strongly the results having from Ref.15, we can conclude that there may be some experimental errors in Ref.1. The calculated electron-phonon coupling constant ($\lambda$) shows contradiction with the other theoretical data but match well with the value (0.32) having from Ref.17 by using the same procedure bearing the reliability of this present calculation. However, using these parameters in Eq. 22 the obtained value of $T_c$ is much lower than that of the experimental and theoretical one but match well with the value (0.01 K) having from Ref.17. This existing discrepancy suggests that the electrons in BaIr$_2$P$_2$ might be strongly coupled for certain phonon modes [17]. In case of BaIr$_2$As$_2$ superconductor the comparison of our estimated superconducting parameters are not possible due to the absence of other theoretical letters. Moreover from Table 7 one can notice that the calculated $\gamma$ match well with the experimental one. The evaluated $T_c$ of BaIr$_2$As$_2$ in this present study is 6 K which is somewhat larger than the experimental one (2.45

K). However, the evaluated $\lambda$ suggests that BaIr$_2$P$_2$ and BaIr$_2$As$_2$ compounds are the weakly coupled and moderately coupled BCS superconductors respectively.

From Fig. 3 (DOS diagram) it is evident that the vibration of Ir-5d and P-3p states in case of BaIr$_2$P$_2$ superconductor and Ir-5d and As-4p states for BaIr$_2$As$_2$ superconductor make large contribution to $\lambda$ as most of the contribution near Fermi level comes from those states [59]. Due to the large contribution of these states at Fermi level the electrons of these states contribute the most for the superconductivity in BaIr$_2$P$_2$ and BaIr$_2$As$_2$ ternary intermetallic compounds.

**IV. Conclusions**

In this present study, we explore the detailed physical properties (structural properties, elastic properties, electronic properties and thermodynamic properties) and superconducting properties of two isostructural bulk superconductors barium iridium phosphide BaIr$_2$P$_2$ ($T_c$ ~ 2.1 K) and barium iridium arsenide BaIr$_2$As$_2$ ($T_c$ ~ 2.45 K) via the first principles method. The calculated structural parameters show reasonably good agreement with the experimental values. DOS and band structure analysis of BaIr$_2$P$_2$ and BaIr$_2$As$_2$ show that both the compounds under consideration are metal. Investigation on the Mulliken atomic populations and total charge density revel that a mixture of ionic, covalent and metallic bonds exist in both the superconductors. Both the compounds show good stability in nature and have anisotropic feature. *B/G* ratio demonstrates ductile behavior of both the superconductors. According to the study of Vickers hardness BaIr$_2$As$_2$ is relatively soft compound compared to that of BaIr$_2$P$_2$. The Debye temperature of both the superconductors is calculated to be 293.06 K and 258.47 K for BaIr$_2$P$_2$ and BaIr$_2$As$_2$ respectively. The study of thermal conductivity revels that at ambient condition both the compounds show comparatively low thermal conductivity. We also explore the melting temperature of both the compounds which indicates that the melting temperature of BaIr$_2$P$_2$ is larger than that of BaIr$_2$As$_2$ superconductor. Moreover, $T_c$ of BaIr$_2$P$_2$ is estimated to be 0.0057 K which is much lower than the experimental value 2.10 K indicating that the electrons in BaIr$_2$P$_2$ might be strongly coupled for certain phonon modes. On the other hand, the evaluated $T_c$ of BaIr$_2$As$_2$ is 6 K shows good consistent with the experimentally evaluated value 2.45 K. However, the evaluated $\lambda$ suggests that BaIr$_2$P$_2$ and BaIr$_2$As$_2$ compounds are the weakly and moderately coupled BCS superconductors respectively.

**Acknowledgements**

I (1st author) would like to thank Assistant Prof. Md. Atikur Rahman (Sir) for giving me the best support during the research period. I would also like to thank all of the honorable teachers of Physics Department, Pabna University of Science and Technology, Bangladesh, for their encouraging speech.

**Author Contributions**

Md.Z. Rahaman conceived the idea and initiated the project. He also performed the detailed calculations and prepared all figures. He also analyzed the data and wrote the manuscript. Md.A. Rahman supervised the project.